\documentclass{ws-ijmpa}
\usepackage[super,compress]{cite}
\usepackage{graphicx}

\usepackage[hidelinks]{hyperref}
\def\ep{\varepsilon}

\begin{document}
\markboth{Maciej Maliborski and Andrzej Rostworowski} {Lecture Notes
  on Turbulent Instability of Anti-de Sitter Spacetime}

%
\catchline{}{}{}{}{}
%

\title{LECTURE NOTES ON TURBULENT INSTABILITY OF ANTI-DE~SITTER SPACETIME
\footnote{This is a written version of the first lecture given by the
  second author at NR/HEP2 Spring School held at IST-Lisbon, 11-14
  March 2013. Accompanying {\sc Mathematica} notebooks are publicly
  available on the School web page\cite{nrhep2}. The second lecture is
  covered in \textit{Phys. Rev. Lett.} \textbf{111}, 051102 (2013),
  \href{http://arxiv.org/abs/1303.3186}{\texttt{arXiv:1303.3186}}.}}

\author{MACIEJ MALIBORSKI}

\address{M. Smoluchowski Institute of Physics, Jagiellonian University, 30-059 Krak\'ow, Poland\\
  maliborski@th.if.uj.edu.pl}

\author{ANDRZEJ ROSTWOROWSKI}

\address{M. Smoluchowski Institute of Physics, Jagiellonian University, 30-059 Krak\'ow, Poland\\
  arostwor@th.if.uj.edu.pl}

\maketitle


\begin{abstract}
In these lecture notes we discuss recently conjectured instability of
anti-de Sitter space, resulting in gravitational collapse of a
large class of \textit{arbitrarily} small initial perturbations. We
uncover the technical details used in the numerical study of
spherically symmetric Einstein-massless scalar field system with
negative cosmological constant that led to the conjectured
instability.

\end{abstract}

\ccode{PACS numbers: 04.25.dc, 04.20.Ex}


\section{Introduction}

Anti-de Sitter (AdS) spacetime is the unique maximally symmetric
solution of the vacuum Einstein equations $G_{\alpha\beta}+\Lambda
g_{\alpha\beta}=0$ with negative cosmological constant $\Lambda$.

Geometrically, AdS$_{d+1}$ can be thought of as being wrapped
infinitely many times around hyperboloid
\begin{equation}
-\left(X^0\right)^2 + \sum _{k=1}^d \left(X^k\right)^2 -
\left(X^{d+1}\right)^2 = -\ell^2,
\end{equation}
embedded in flat, $O(d,2)$ invariant space, with a line element
\begin{equation}
ds^2 = -\left(dX^0\right)^2 + \sum _{k=1}^d \left(dX^k\right)^2 -
\left(dX^{d+1}\right)^2.
\end{equation}
Parametrization
\begin{equation}
X^0 = \ell\, \sec x \, \cos t, \quad X^{d+1} = \ell\, \sec x \, \sin
t, \quad X^k = \ell\, \tan x\, n^k,
\end{equation}
with
\begin{equation}
-\infty < t < +\infty, \quad 0 \leq x < \pi/2, \quad \sum _{k=1}^d
\left( n^k \right)^2 =1,
\end{equation}
yields induced metric on the hyperboloid in the form
\begin{equation}
  ds^2 = \frac {\ell^2}{\cos^{2} x} \left( -dt^2 + dx^2 +  \sin^{2} x\, d\Omega^2_{S^{d-1}} \right).
\end{equation}
This metric is indeed a solution to the vacuum Einstein equations with
\mbox{$\Lambda=-d(d-1)/(2\ell^2)$}. Conformal infinity for this metric,
located at $x=\pi/2$ is the timelike cylinder
$\mathcal{I}=\mathbb{R}\times S^{d-1}$ with the boundary metric
$ds^2_{\mathcal{I}}=-dt^2 + \sin^{2} x \, d\Omega^2_{S^{d-1}}$.

Asymptotically anti-de Sitter (AAdS) spacetimes (that is spacetimes
which share the conformal boundary with AdS but may be very different
in the bulk, in particular may contain horizons) have come to play a
central role in theoretical physics, prominently due to the AdS/CFT
correspondence which conjectures a duality between gravity in the AdS
bulk and a quantum conformal field theory with a large number of
strongly interacting degrees of freedom living in the spacetime
corresponding to the AdS conformal boundary\cite{mal,wit}. By the
positive energy theorem, AdS spacetime is a ground state among AAdS
spacetimes, much as Minkowski spacetime is a ground state among
asymptotically flat spacetimes. However, the evolution of small
perturbations of these ground states are different. In the case of
Minkowski, small perturbations disperse to infinity and the spacetime
is asymptotically stable \cite{ck}. In contrast, asymptotic stability
of AdS is precluded because the conformal boundary acts like a mirror
at which perturbations propagating outwards bounce off and return to
the bulk that results in complex nonlinear wave interactions in an
effectively bounded domain. Understanding of these interactions is the
key to the problem of stability of AdS spacetime.

A recent numerical and analytic study of the four dimensional
spherically symmetric Einstein-massless scalar field equations with
negative cosmological constant indicated that AdS space may be
unstable against the formation of a black hole under large class of
arbitrarily small perturbations \cite{br}. Qualitatively the same
results were obtained later in higher dimensions \cite{jrb,
  bll}. Although gravitational collapse seems to be a generic fate of
a small perturbation of AdS, it was suggested in \cite{br} that there
may also exist a set of initial data for which the evolution remains
globally regular in time. This conjecture was substantiated in
\cite{mr}, where the evidence for the existence of globally regular,
nonlinearly stable, time-periodic solutions in the Einstein AdS -
massless scalar field system was given. Similar class of globally
regular, nonlinearly stable, asymptotically AdS solutions, namely
boson stars was studied in \cite{bll_BS}. The similar behavior is
expected for the pure vacuum case with simplifying symmetry
assumptions \cite{mr_prep} and with no symmetry assumptions \cite{dhs,
  dhms}. The outcome of all these studies suggests that the structure
of phase space for gravity with negative cosmological constant is
complicated and probably for still some time numerical simulations
will be the key tool to investigate this problem and to assist
analytic attempts like \cite{im}. Thus in this short lecture notes we
will concentrate on numerical tricks and details used in numerical
codes of \cite{br,jrb} that are usually omitted in research papers,
however crucial they are for stable long-time numerical simulations of
AAdS spacetimes.

\section{The model}

To make the long-time numerical investigation of AdS stability
tractable we assume spherical symmetry. Since by Birkhoff's theorem
spherically symmetric vacuum solutions are static, we need to add
matter to generate dynamics. A simple matter model is the minimally
coupled massless scalar field in $d+1$ spacetime dimensions
\cite{br,jrb}:
\begin{align} & G_{\alpha\beta}+\Lambda g_{\alpha\beta} = 8 \pi G
  \left(\partial_{\alpha} \phi \,\partial_{\beta} \phi - \frac{1}{2}
    g_{\alpha\beta} \partial_\mu \phi \partial^\mu \phi \right)\,, \\
  & g^{\alpha\beta} \nabla_{\alpha} \nabla_{\beta} \phi=0 \, .
\end{align}
Let us recall that in the asymptotically flat case ($\Lambda=0$) this
model has led to important insights, such as the proof of the weak
cosmic censorship by Christodoulou \cite{ch1,ch2} and the discovery of
critical phenomena at the threshold for black hole formation by
Choptuik \cite{matt}. We parametrize the $(d+1)$--dimensional
asymptotically AdS metric by the ansatz
\begin{equation}
  \label{adsd+1:ansatz}
  ds^2\! =\! \frac {\ell^2}{\cos^2{\!x}}\left( -A e^{-2 \delta} dt^2 + A^{-1} dx^2 + \sin^2{\!x} \,  d\Omega^2_{d-1}\right),
\end{equation}
where $\ell^2=-d(d-1)/(2\Lambda)$, $d\Omega^2_{d-1}$ is the round
metric on $S^{d-1}$, $-\infty<t<\infty$, $0\leq x<\pi/2$, and
$A$, $\delta$ are functions of $(t,x)$. For this ansatz the evolution
of a self-gravitating massless scalar field $\phi(t,x)$ is governed
(see the accompanying {\sc Mathematica} notebook
\texttt{equations.nb} \cite{nrhep2}) by the following system (using
units in which $8\pi G=d-1$)
\begin{align}
  \label{ms_in_ads_d+1:eq_wave}
  \dot\Phi & = \left( A e^{-\delta} \Pi \right)',
  \quad \dot \Pi = \frac{1}{\tan^{d-1}{\!x}}\left(\tan^{d-1}{\!x} \,A e^{-\delta} \Phi \right)',\\[1ex]
  \label{ms_in_ads_d+1:eq_00}
  A' \!&= \!\frac{d-2+2\sin^2{\!x}} {\sin{x}\cos{x}} \, (1-A) -
  \sin{x}\cos{x} \, A \left( \Phi^2 + \Pi^2 \right),
  \\[1ex]
  \label{ms_in_ads_d+1:eqs_10_11}
  \delta' \!&=\! - \sin{x}\cos{x} \left( \Phi^2 + \Pi^2 \right),
\end{align}
where ${}^{\cdot}=\partial_t$, ${}'=\partial_x$, and
\begin{equation}
  \label{Phi_Pi_definitions}
  \Phi= \phi', \qquad \Pi= A^{-1} e^{\delta} \dot \phi \,.
\end{equation}
Note that the length scale $\ell$ drops out from the equations.  This
system has a one-parameter family of static solutions ($\phi=0$,
$\delta=\mbox{const}$, $A = 1 - M \cos^{2} x / \left( \tan{x}
\right)^{d-2}$) which are Schwarzschild-AdS black holes for $M>0$ and
the pure AdS for $M=0$.  In analogy to the Schwarzschild-AdS black
hole the mass function for the system can be defined as $m(t,x) =
\left(1-A(t,x)\right) \, \sec^{2} x \, \left( \tan x \right)^{d-2}$.
We restrict our attention to smooth solutions with a finite mass
\begin{equation}
  \label{mass}
  M(t):=\lim_{x \rightarrow \pi /2} m(t,x) = \int\limits_0^{\pi/2}
  \left(A\Phi^2+A\Pi^2\right) \left(\tan x\right)^{d-1} \,dx < \infty
  \,.
\end{equation}
Smoothness at spatial infinity and the requirement of the total mass
(\ref{mass}) to be finite, imply that near $x = \pi/2$ (using $z =
\pi/2 - x$) (see the accompanying {\sc Mathematica} notebook
\texttt{boundary.nb} \cite{nrhep2})
\begin{align}
  A(t,x) &= 1 - M \, z^d + \mathcal{O}\left(z^{d+2} \right),
  \quad \delta(t,x) = \delta_{0}(t) +
  \mathcal{O}\left(z^{2d}\right), \nonumber
  \\
  \phi(t,x) &= f_{0} + f_d(t)\, z^d + \mathcal{O} \left(z^{d+2}\right),
  \label{phi_boundary}
\end{align}
where the subsequent terms in the expansions are expressed by constant
$M$ and the functions $f_d(t)$, $\delta_{0}(t)$ and their
derivatives. These are in turn uniquely determined by the evolution of
initial data \footnote{In particular, for compactly supported initial
  data $f_{0}=0$.}.  Thus in this particular model if we require the
total mass to be finite it implies the mass to be conserved as
well. The local well-posedness of the above initial-boundary value
problem was proved in \cite{hs1}.  It is important to stress that for
\textit{odd} $d$ higher even terms in the expansion for $\phi$ do not
vanish (they vanish identically \textit{only} in the limit $M
\rightarrow 0$). \textbf{Exercise~1:} verify this statement using
\texttt{boundary.nb} notebook. Thus for $odd$ $d$ the boundary
behavior of the scalar field is not compatible with eigenfunctions of
the linear self-adjoint operator, that governs the evolution of a
massless scalar field on a \textit{fixed} AdS${}_{d+1}$
background. This operator reads $L = -(\tan x)^{1-d} \partial_x
\left((\tan x)^{d-1} \partial_x \right)$. Its spectrum is given by
$\omega_j^2=(d+2j)^2$, $j=0,1,...\,$ and the eigenfunctions read
\begin{equation}
  \label{eigen_modes}
  e_j(x) = 2 \frac {\sqrt{j!\, (j+d-1)!}} {\Gamma\left(j + d/2\right)} \cos^{d} x \, P_j^{(d/2-1,d/2)} (\cos 2x)\,,
\end{equation}
where $P_j^{(\alpha,\beta)} (x)$ are the Jacobi polynomials. These
eigenfunctions form an orthonormal base in the Hilbert space of
functions $L^2\left([0,\pi/2], \, (\tan x)^{d-1} \, dx\right)$. Below
we denote the inner product on this Hilbert space by $\left( f \left|
g \right. \right) := \int _0 ^{\pi/2} f(x) g(x) (\tan x)^{d-1}\, dx$.

\section{Numerical method}

We solve the system
(\ref{ms_in_ads_d+1:eq_wave})-(\ref{ms_in_ads_d+1:eqs_10_11})
numerically using a fourth-order accurate finite-difference
approximation for spatial derivatives in evolution equations. For the
general introduction to finite difference approximation see Hirotada
Okawa's lecture in this volume. To integrate the evolution equations
in time we use the method of lines and a 4th-order Runge-Kutta (RK4)
scheme, where at each time step the metric functions are updated by
solving constraint equations: the Hamiltonian constraint
(\ref{ms_in_ads_d+1:eq_00}) and the slicing condition
(\ref{ms_in_ads_d+1:eqs_10_11}) with RK4 in space. The (adjustable)
time step $\Delta t$ is kept $1/6 \leq e^{-\delta(t,\pi/2)} \Delta t /
\Delta x \leq 1/3$ for the constant spatial grid spacing $\Delta
x$. Preservation of the momentum constraint $\dot A + 2 \sin{x}\cos{x}
\, A^2 e^{-\delta} \Phi\, \Pi =0$ can be monitored to check the
accuracy of the code. This (fully constrained) scheme allows for
stable long-time evolution. Another crucial ingredient of stable
evolution of nonlinear evolution equations on compact domains is
careful treatment of boundaries of numerical domain. Here as method of
thumb we avoid nonsymmetric stencils at the boundary. Instead we try
to put as much analytic information as we have into the code. That is
we use:
\begin{enumerate}
\item Symmetry of the function. If we know that the function is
  symmetric or antisymmetric with respect to the boundary of
  numerical domain we use this information in finite-difference
  approximation for the derivatives of this function at the
  boundary. For example in the 4th order finite difference
  approximation for the 1st derivative of the function $f(x)$ at $x=0$
\begin{equation}
  f'(0) = (f(-2\Delta x) - 8f(-\Delta x) + 8f(\Delta x) - f(2\Delta x))/(12 \Delta x) + \mathcal{O}\left(\Delta x^4\right)
\end{equation}
we can put $f(-k \, \Delta x) = \pm f(k \, \Delta x)$ for symmetric
and antisymmetric functions respectively.
\item Taylor expansion of the function. If we know the Taylor
  expansion of the function at the boundary of numerical domain we can
  use this information to express the function at the boundary with
  its values inside the domain. For example if the function $f(x)$ has
  the following Taylor expansion at $x=0$ $f(x) = f_0 + f_2 x^2 + f_4
  x^4 + \mathcal{O}\left(x^6\right)$, then from the set of linear
  equations $f(k \, \Delta x) = f_0 + f_2 (k \, \Delta x)^2 + f_4 (k
  \, \Delta x)^4$ for $k=1,2,3$ we can get the coefficients $f_0$,
  $f_2$, $f_4$ and in particular $f(0) = f_0$.
\item De l'H\^ospital's rule. It often happens that the right hand
  side of evolution equations at the boundary is of the form $0/0$ (as
  happens here for $\dot\Pi$ in (\ref{ms_in_ads_d+1:eq_wave}) at $x=0$
  and $x=\pi/2$). A useful numerical trick to deal in such case is to
  use de l'H\^ospital's rule. For example suppose that
  $f(x)\stackrel{x \rightarrow 0}{\longrightarrow} 0$. Then to deal
  with numerical instabilities produced by $f(x)/x$ at small $x$ one
  can use the identity $f/x = f'(x) - x (f/x)'$, where the
  instabilities coming from small denominators are suppressed by the
  small factor $x$ comparing to the leading term $f'(x)$. We have
  learned this useful trick from An\i l Zengino\u{g}lu during his stay
  in Cracow in 2010.
\end{enumerate}
In particular, in our code numerical boundaries ($x=0$ and $x=\pi/2$)
are treated as follows. We put Eq.~(\ref{ms_in_ads_d+1:eq_wave}) for
$\dot \Pi$ in the form
\begin{equation}
\label{ms_in_ads_d+1:eq_wave(2)}
\dot \Pi = \left(A e^{-\delta} \Phi \right)' + (d-1)\frac{A e^{-\delta} \Phi}{\sin{\!x} \cos{\!x}}\,,
\end{equation}
where close to the boundaries we use de l'H\^ospital's rule:
\begin{equation}
\label{Hospital}
\frac{A e^{-\delta} \Phi}{\sin{\!x} \cos{\!x}} = \frac{\left(A
    e^{-\delta} \Phi \right)'}{\cos{\!2x}} -\frac{1}{2} \tan{\!2x}
\left(\frac{A e^{-\delta} \Phi}{\sin{\!x} \cos{\!x}} \right)' \,.
\end{equation}
This is crucial for the stable evolution at $x=\pi/2$ and convenient in
the center $x=0$: if de l'H\^ospital rule (\ref{Hospital}) is used in
the center, numerical dissipation can be turned off. If the de
l'H\^ospital trick is not applied at $x=0$ then one should use the
equation for $\dot \Pi$ in the form (\ref{ms_in_ads_d+1:eq_wave}), as
the form (\ref{ms_in_ads_d+1:eq_wave(2)}) is numerically unstable(!)
in the symmetry center. The values $\Phi(t,0)$, $\Phi(t,\pi/2)$,
$\Pi(t,\pi/2)$ are set to zero due to boundary behavior. The values
$\Pi(t,0)$, $\Phi(t,\pi/2-\Delta x)$, $\Pi(t,\pi/2-\Delta x)$ are set
from the Taylor expansion at the boundaries. Symmetry properties of
$\Phi$, $\Pi$, $A$ and $\delta$ in the center $x=0$ are used when
needed in evaluating spatial derivatives on the right hand side of evolution
equations near the center. In the on-line material \cite{nrhep2} we
include the {\sc Mathematica} notebook \texttt{evolution.nb} that
allows for testing of what was stated above \footnote{In the
  \texttt{evolution.nb} notebook, for the clarity of the code, we use
  nonsymmetric stencils for evaluating $\dot \Phi(t,\pi/2-\Delta x)$,
  $\dot \Pi(t,\pi/2-\Delta x)$ instead of setting $\Phi(t,\pi/2-\Delta
  x)$, $\Pi(t,\pi/2-\Delta x)$ from the Taylor expansion at the
  boundary.}.  \textbf{Exercise~2}: experiment with
\texttt{evolution.nb} switching de l'H\^ospital trick at $x=\pi/2$
and/or $x=0$ on and off; see what happens.  \textbf{Exercise~3}: turn
de l'H\^ospital trick at $x=0$ off and implement the form
(\ref{ms_in_ads_d+1:eq_wave(2)}) in the center; see what happens.

At each time step we check for the minimum of $A(t,x)$. If it drops
below a certain threshold (that we set to be $2^{-k+7}$ on the grid
with $n=2^k+1$ points, with $k\geq10$) we say that an apparent horizon
forms and we stop the evolution.

If transfer of energy to higher modes of linearized problem is present
during the evolution of initial data, then higher spatial resolution
may be needed. If that happens we perform global mesh refinement from
$\Delta x$ to $\Delta x/2$ grid spacing. This brute force approach
requires a lot of computational power and should be replaced with some
more subtle algorithm in the future. At this point the brute force is
supplied with parallel computing. We use MPI library (for a concise
introduction to MPI programming see \cite{MPI}). While stepping in
time in evolution equations (\ref{ms_in_ads_d+1:eq_wave}) is naturally
suited for parallelization, solving the constraints
(\ref{ms_in_ads_d+1:eq_00}, \ref{ms_in_ads_d+1:eqs_10_11}) can be a
thin throat for parallel computing. We solve this problem in the way,
that processors are not synchronized on the grid. That is, while
solving a constraint equation (what is being done in sequence from
$x=0$ to $x=\pi/2$) a processor waits only for its nearest left-hand
neighbor on the grid to complete the task. As a result processors are
not synchronized and these starting solving the constraints (at $x=0$
in our case) can be a few steps in time ahead comparing to those
finishing a given time step (on a part of the grid close to $x=\pi/2$
in our case). In this setting, time step and horizon formation control
that requires exchange of information between non-neighboring
processors is realized with non-blocking MPI library procedures
\texttt{MPI\_isend} and \texttt{MPI\_irecv}. In this way we get a
scaling with the factor of two with the number of cores.

\section{Numerical results and their origin}

\begin{figure}[h]
  \includegraphics[width=0.48\textwidth]{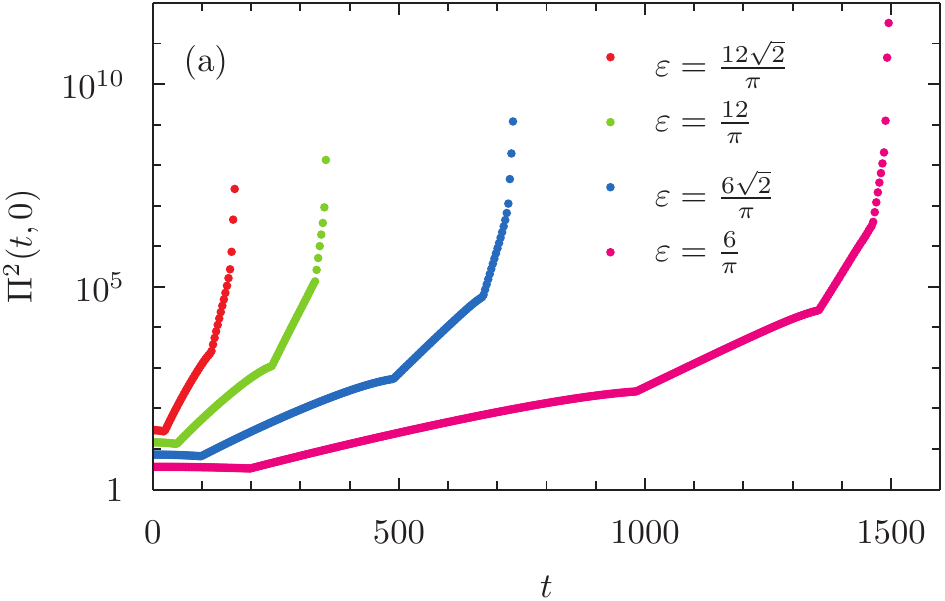}
  \hspace{2.2ex}
  \includegraphics[width=0.48\textwidth]{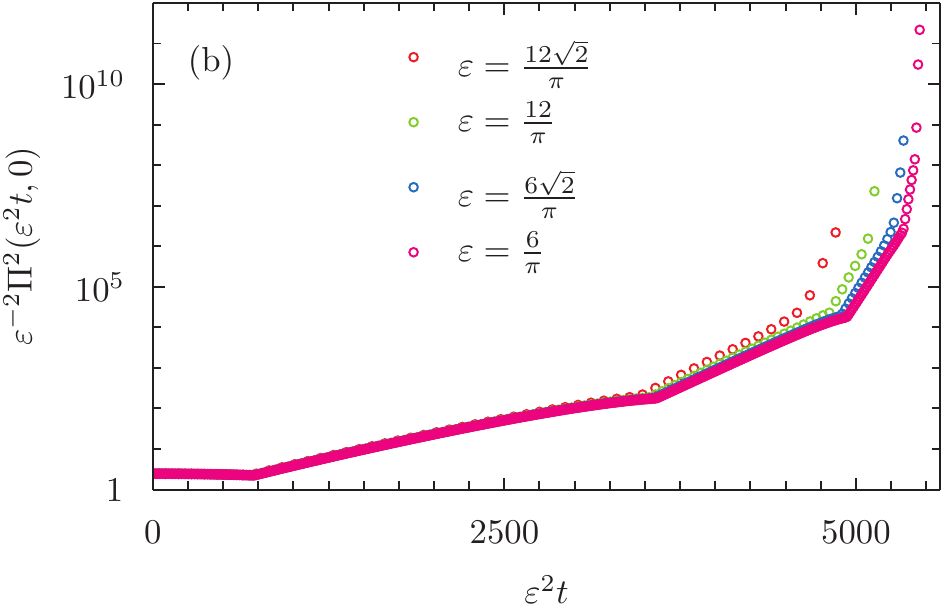}
  \caption{(a) $\Pi^2(t,0)$ for solutions with initial data
    \eqref{idata} for three moderately small amplitudes.  For clarity
    of the plot only the upper envelopes of rapid oscillations are
    depicted.  After making between about fifty (for
    $\ep=12\sqrt{2}/\pi$) and five-hundred (for $\ep=6/\pi$)
    reflections, all solutions finally collapse. (b) The curves from
    the plot (a) after rescaling $\ep^{-2} \Pi^2(\ep^2 t,0)$.}
  \label{fig1}
\end{figure}

Our key numerical argument for AdS instability is depicted in
Fig.~\ref{fig1} (quoted from \cite{br}). We numerically evolve initial
data of the form
\begin{equation}
\label{idata}
\Phi(0,x)=0\,,\quad \Pi(0,x)=\ep \exp\left(-\frac{4 \tan^2{\!x}}{\pi^2\sigma^2}\right)\,,
\end{equation}
with fixed width $\sigma=1/16$ and varying amplitude $\ep$. For such
data the scalar field is well localized in space and propagates in
time as a narrow wave packet. We look at the Ricci curvature in the
center, $\ell^2\,R(t,0)=-2\Pi^2(t,0)-12$, that can serve as a
good indicator for the onset of instability. This quantity oscillates
with frequency $\approx 2$ (as it takes time $\approx \pi$ for the
wave packet to make the round trip from and back to the center). An
upper envelope of the oscillations of $\Pi^2(t,0)$ is shown in
Fig.~1a. We see that after some time, when the amplitude remains
approximately constant, there begins a subsequent phases of (roughly)
exponential growth, until finally the solution collapses. It can be
seen that the time of onset of the subsequent phases scales as
$\ep^{-2}$ (see Fig.~1b). Indeed, after the rescaling $\Pi^2(t,0)
\rightarrow \ep^{-2}\Pi^2(\ep^{2}t,0)$ all curves of Fig.~1a seem to
converge to some universal curve (for this family of initial data with
$\sigma$ fixed to $1/16$), that means that arbitrarily small
perturbations eventually start growing. Note that this behavior is
morally tantamount to instability of AdS space, regardless of what
happens later, in particular whether the solution will collapse or
not. In the second part of \cite{br} we sketched the mechanism behind
this numerically discovered instability of AdS in the framework of
weakly nonlinear perturbation theory (that we previously extensively
used to predict nonlinear tails of dispersing solution in
asymptotically flat case). To summarize there are two key ingredients
to the mechanism of the instability of AdS:
\begin{enumerate}
\item \textbf{Conservation of the total mass of the system}. The first
  crucial ingredient is conservation of the total mass of the system,
  that (opposite to a flat case) can not be dispersed to
  infinity. Conformal structure of AdS space makes it effectively
  bounded and to make the initial-boundary value problem well posed
  setting boundary conditions may be needed. In the model at hand
  reflecting boundary conditions at conformal infinity follow directly
  from the regularity of solutions of Einstein equations at the
  boundary and the requirement of the total mass (\ref{mass}) to be
  finite. More general models, like conformally coupled scalar field
  or AdS-Einstein-Yang-Mills system, allow for more freedom in setting
  boundary conditions, where the conservation of the total mass at
  conformal infinity can be picked up as physically natural boundary
  condition. However one should keep in mind that other choices are
  possible as well. We are grateful to Helmut Friedrich for stressing
  this point.
\item \textbf{Resonant spectrum}. The second crucial ingredient for
  instability is the resonant spectrum of the wave operator on the
  \textit{fixed} AdS background (see the previous section). That is
  the spectrum consists of equidistantly spaced real numbers. Once the
  modes of the corresponding wave operator are coupled through gravity
  (or nonlinearity) it results in resonant coupling between the modes
  that generically, efficiently transfers (conserved) energy into
  higher and higher modes. We call this process a turbulent transfer
  of energy, in analogy to Navier-Stokes equations that transfer the
  eddies to smaller and smaller spatial scales. In the case of
  Einstein equations this process is ultimately cut by a black hole
  formation, but mind the special case of $d=2$
  \cite{bj}. Interestingly, it seems that \textit{asymptotically}
  resonant spectrum is also good enough for such turbulent transfer of
  energy to take place, see \cite{m}.
\end{enumerate}
To demonstrate the transfer of energy to higher frequencies we define
the Fourier coefficients $\Phi_j:=\left( e'_j \left| \sqrt{A}\,\Phi
\right. \right)$ and $\Pi_j:=\left( e_j \left| \sqrt{A}\,\Pi
\right. \right)$ and express the total (conserved) mass as the
Parseval sum $M=\sum_{j=0}^{\infty} E_j(t)$, where
$E_j:=\Pi_j^2+\omega_j^{-2} \Phi_j^2$ is the $j$-mode energy.  The
evolution of the energy spectrum, that is the distribution of mass
among the modes, is depicted in Fig.~2 for gaussian initial data
\eqref{idata} with $\ep=6/\pi$. Initially, the energy is concentrated
in low modes; the exponential cutoff of the spectrum expresses the
smoothness of initial data. During the evolution the range of excited
modes increases and the spectrum becomes broader. Just before horizon
formation the spectrum exhibits the power-law scaling $E_{j}\sim
j^{-\alpha}$ with exponent $\alpha\approx 6/5$. This value seems to be
universal, i.e., the same for all initial data, but it changes with
dimension~$d$. Our preliminary guess based on studying the instability
in $d=3,4,5$ is that
\begin{equation}
\label{guess}
\alpha(d) = 6/5 + 4(d-3)/5\,.
\end{equation}
Note that the formation of a black hole provides a cutoff for the
turbulent energy cascade (in amusing analogy to viscosity for the
turbulent cascade in fluids).  Clearly, the formation of the power-law
spectrum reflects the loss of smoothness of the solution during
collapse; it would be very interesting to compute $\alpha(d)$
analytically.

The plot in Fig.~2 was made for the instants of time when the scalar
field was imploding through the center, and decaying exponentially
fast at infinity (because we started with compactly supported initial
data). Mind that, in $odd$ number of spatial dimensions $d$, if we
expressed the total mass as the Parseval sum over the modes at
instants of time when the scalar pulse is concentrated close to
infinity (with a polynomial decay according to \eqref{phi_boundary})
we would get steep but still polynomial tail of energy spectrum just
because of the fact that the basis of the eigenfunctions of the wave
operator on a \textit{fixed} AdS background is incompatible (in
\textit{odd} d) with the boundary behavior of the massless scalar
field in the self-gravitating case. Of course, such polynomial decay
would have nothing to do with the effect depicted in Fig.~2 and would
just illustrate the fact that the $e_j(x)$ base is not well adapted to
self-gravitating case in \textit{odd} number of spatial dimensions.

\begin{figure}[h]
    \centerline{\includegraphics[width=0.70\linewidth]{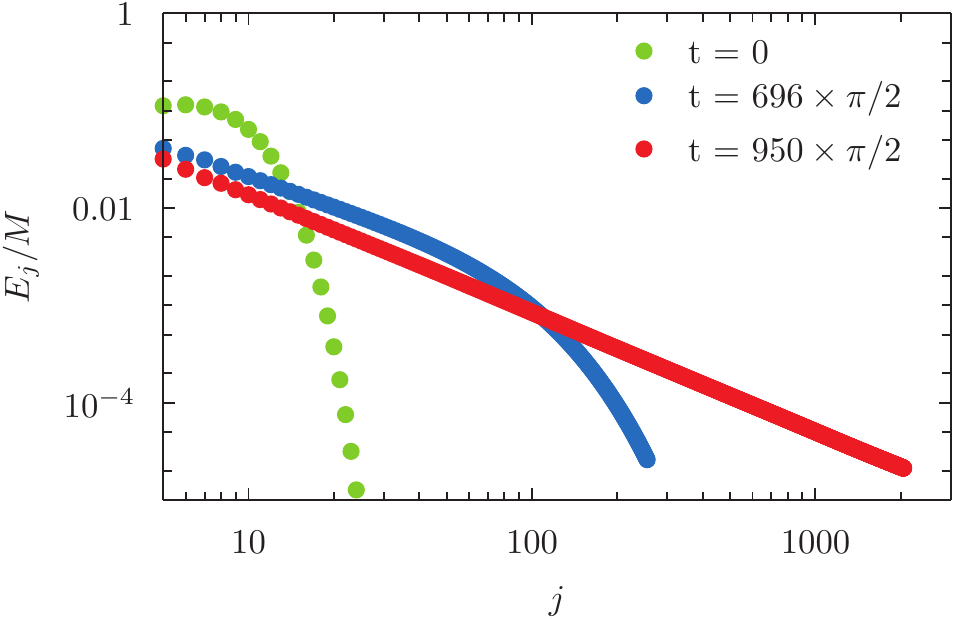}}
    \caption{Log-log plot of the energy spectrum at three moments of
      time: initial, intermediate, and just before collapse for the
      solution evolving from gaussian initial data \eqref{idata} with
      $\ep=6/\pi$. The fit of the power law $E_{j}\sim j^{-\alpha}$ at
      time $t=950\times\pi/2$ gives the slope $\alpha\approx 6/5$.}
\end{figure}

\section{Time-periodic solutions}

This section is in fact only a short overview of the paper \cite{mr},
covering the second lecture delivered by one of the authors at NR/HEP2
school \cite{nrhep2}.

However weakly nonlinear perturbation theory provides a good hint for
explanation of the mechanism of AdS instability it fails to provide
quantitative predictions for a given initial data, like time of the
black hole formation, its initial radius, \textit{etc.} Still it
proved to be extremely useful in perturbative construction of
time-periodic solutions.

It turns out that negative cosmological constant allows for stable,
globally regular time-periodic solutions \cite{mr}. In that paper we
constructed time-periodic solutions with two independent
methods. Perturbative method consists in representing the
time-periodic solution as a perturbative series in the amplitude of a
mode dominating perturbative solution. The subsequent terms in this
expansion are determined by removing resonant terms that generically
arise. Numerical construction is build upon new spectral code well
suited to study the solutions that do not collapse. Convergence of
results of these two independent methods makes us feel confident in
the results. Moreover numerical evolution of the initial data for a
time-periodic solutions found with those methods reproduces this
solution. This is a strong evidence that those time-periodic solutions
are nonlinearly stable.

After these lectures had been delivered it was shown in\cite{bll_BS},
that the initial data (\ref{idata}) with $\sigma\approx0.5$ are immune
to the turbulent instability. We are building-up the evidence
\cite{mr2} that the explanation to this immunity comes from the
existence of nonlinearly stable time-periodic solutions in the
system. Namely, for the initial data (\ref{idata}) there is a window
in $\sigma$, where they belong to the stability island of
time-periodic solution bifurcating from the $e_0(x)$ mode of the
massless scalar field propagating on a \textit{fixed} AdS${}_{3+1}$
background.

\section{Summary and outlook}

In these lecture notes we discussed recently conjectured instability
of anti-de Sitter space, highlighting numerical techniques used in the
study of spherically symmetric Einstein - massless scalar field system
with negative cosmological constant\cite{br,jrb}. This case is a neat
example of the fact that from numerical explorations of Einstein's
equations there can grow understanding, conjectures, roads to proofs
and phenomena that would not have been imaginable in the pre-computer
era. The role of computation in general relativity seems destined to
expand in future, playing a role of a telescope for a theoretical
physicist.

The dynamics of AAdS turns out to be an exceptional
meeting point of fundamental problems in general relativity,
PDE theory, theory of turbulence, and high energy
physics. Understanding of these connections is still at its infancy.
The issues discussed here raise probably more questions than give
answers.
\begin{itemize} 
\item The proofs of instability of AdS and the existence of time-periodic
  solutions are still lacking. There is a natural question here: what
  makes the AdS spacetime unstable, while making some other AAdS
  solutions\cite{mr,bll_BS,dhms} stable. Probably it would be
  rewarding to understand the resonant coupling between the modes and
  the energy spectrum \eqref{guess} in more detail. There is also
  some tension between numerical\cite{m} and heuristic
  results\cite{dhms} to what extent only \textit{asymptotically}
  (that is for high wave numbers) resonant spectrum is good enough for
  a turbulent transfer of energy to take place.
\item So far, our studies have been restricted to spherical symmetry
  (or other effectively $1+1$ dimensional settings). Thus relaxing
  symmetry assumptions should be the next step. The heuristic
  arguments\cite{dhs,dhms} suggest that the similar picture of
  instability, with some basins of attraction of some stable AAdS
  solutions emerges also outside spherical symmetry.  The numerical
  studies of axially symmetric AAdS spacetimes containing a black hole
  have been initiated in\cite{bpg} but to build a code capable of
  evolving small perturbations of AdS and detecting eventual black
  hole formation there is still some way to go.
\item The next question (closely related to the previous point) is
  about the end state of evolution. The AdS-Schwarzschild solution,
  being stable \cite{hs2}, is a natural candidate in spherical
  symmetry. However outside spherical symmetry, the problem is more
  delicate as small AdS-Kerr black holes are expected to be unstable
  due to superradiant instability\cite{cd,hr}
\item There are some more general models like conformaly coupled
  scalar field or Einstein-Yang-Mills system, allowing for more
  freedom in prescribing boundary conditions at the timelike boundary of
  AAdS. We are currently studying these more general boundary
  conditions.
\item It might be rewarding to understand better AdS/CFT counterpart
  of stable AAdS solutions.
\end{itemize}
Hopefully we will see some advances at least on the part of these
problems in not too far future.

\section*{Acknowledgments}

It is a pleasure to thank the organizers of the NR/HEP2 Spring School
for giving the possibility to deliver these lectures and for the
stimulating environment for the second lecture to be completed in
time. We are indebted to Piotr Bizo\'n for fruitful collaboration in
investigating the problem of AdS instability. A.R. is particularly
grateful to Tadeusz~Chmaj for teaching the bases of numerical methods
of solving evolution PDEs during long term collaboration, and to
Joanna~Ja\l mu\.zna for the introduction to the use of MPI
library. This work was supported by the NCN grant
DEC-2012/06/A/ST2/00397.

\section*{Appendix. List of publicly available codes}
There are three publicly available {\sc Mathematica} notebooks, to be
downloaded from NR/HEP2 Spring School web page\cite{nrhep2} that are
integral part of these lecture notes:
\begin{itemize}
\item \texttt{equations.nb} - derivation of the system of equations (\ref{ms_in_ads_d+1:eq_wave})-(\ref{ms_in_ads_d+1:eqs_10_11}).
\item \texttt{boundary.nb} - derivation of the boundary behavior (\ref{phi_boundary}).
\item \texttt{evolution.nb} - schematic code for time evolution.
\end{itemize}

\end{document}